\documentclass[aps,pre,preprint, superscriptaddress]{revtex4}
\usepackage{latexsym}
\usepackage{natbib}
\usepackage{amsmath}
\usepackage{amsfonts}
\usepackage{amssymb}
\usepackage[utf8]{inputenc}
\usepackage[T1]{fontenc}
\usepackage[dvips]{color,graphicx}   
\usepackage{verbatim}   
\usepackage{t1enc}
\usepackage{anysize}
\usepackage{tocbibind}
\usepackage{subfig}
\usepackage[percent]{overpic} 
\usepackage{float}  
\usepackage{appendix} 
\usepackage{booktabs} 
\usepackage{natbib}
\usepackage{array}

\begin{document}
\title{Emergent networks in fractional percolation} \author{Lucas
  D. Valdez} \affiliation{Instituto de
  Investigaciones F\'isicas de Mar del Plata (IFIMAR)-Departamento de
  F\'isica, FCEyN, Universidad Nacional de Mar del Plata-CONICET, Mar
  del Plata 7600, Argentina.}
\author{L. A. Braunstein}\affiliation{Instituto de Investigaciones
  F\'isicas de Mar del Plata (IFIMAR)-Departamento de F\'isica, FCEyN,
  Universidad Nacional de Mar del Plata-CONICET, Mar del Plata 7600,
  Argentina.} \affiliation{Physics Department, Boston University,
  Boston, MA 02215, United States} \date{\today}

\begin{abstract}
Real networks are vulnerable to random failures and
malicious attacks. However, when a node is harmed or damaged, it may remain partially
functional, which helps to maintain the overall network structure and
functionality. In this paper, we study the network structure for a
fractional percolation process [Shang, Phys. Rev. E 89, 012813
  (2014)], in which the state of a node can be either fully functional
(FF), partially functional (PF), or dysfunctional (D). We develop new
equations to calculate the relative size of the percolating cluster of
FF and PF nodes, that are in agreement with our stochastic
simulations. In addition, we find a regime in which the percolating cluster can
be described as a coarse-grained bipartite network, namely, as a set of
finite groups of FF nodes connected by PF nodes. Moreover, these
groups behave as a set of ``supernodes'' with a power-law degree
distribution. Finally, we show how this emergent structure explains
the values of several critical exponents around the percolation
threshold.

\end{abstract}

\maketitle

\section{Introduction}
Real-world networks, such as social and infrastructure networks, are
continuously facing natural and man-made threats that compromise their
structure and
functionality~\cite{Vespignani2010,garnett2020vulnerability}. For
instance, extreme flooding in urban areas may lead to extensive damage
to infrastructures and promote the spread of water-borne and
vector-borne
diseases~\cite{mejia2021flood,chen2012effects}. Similarly, the
collapse of an electrical transmission tower, due to poor maintenance,
extreme weather, or a malicious attack, could trigger a cascading
failure in the power
grid~\cite{chen2009review,sullivan2017cyber,Exxon01}. Because such
networks are constantly exposed to failures, many researchers have
focused their attention on understanding how damage affects network
structure and functionality. In particular, percolation
theory~\cite{stauffer2014introduction,stauffer1979scaling,saberi2015recent,li2021percolation}
and network science~\cite{Newman-2006,barabasi2013network} have been
used extensively in these investigations, as they provide tools for
assessing the robustness of networks to
failures~\cite{li2015network,mureddu2016islanding,li2021percolation,li2019recent,valdez2020cascading}.

One of the simplest models to study damaged systems in percolation
theory is node
percolation~\cite{stauffer2014introduction,stauffer1979scaling,li2021percolation},
in which a fraction $1-p$ of nodes are randomly and independently
removed (failed or vacant), while the remaining fraction $p$ of nodes
are intact (occupied). Here, $p$ is also called the control
parameter. In this model, several quantities of interest are studied,
such as: 1) the distribution of finite cluster sizes $n_s$ where $s$
denotes the cluster size, 2) the mean finite cluster size $\langle s
\rangle$, and 3) the probability $P_{\infty}$ that a randomly chosen
node belongs to a cluster with a macroscopic size, called the
percolating cluster or the giant component (GC).  It was shown that in
node percolation, a second-order phase transition occurs at a critical
threshold $p=p_c$ when the total number of nodes $N$ tends to
infinity. Above this threshold, i.e., in the percolating phase, a GC
emerges ($P_{\infty}>0$), whereas below $p_c$, the system is composed
solely of finite clusters ($P_{\infty}=0$), which is called the
non-percolating phase. In addition, around $p\approx p_c$, several
magnitudes behave as power-laws, such as $P_{\infty}\sim
(p-p_c)^{\beta}$ for $p\gtrsim p_c$, $\langle s \rangle \sim
|p-p_c|^{-\gamma}$ (i.e., the mean finite cluster size diverges at the
critical point), and $n_s\sim s^{-\tau}$ at $p=p_c$, where $\beta$,
$\gamma$, and $\tau$ are called critical
exponents~\cite{stauffer2014introduction,stauffer1979scaling}.  The
value of $p_c$ for $N\to \infty$ depends on the network structure or
topology. On the contrary, the values of the critical exponents do not
depend on the specific structure but rather on the network dimensionality and
the node
degree-heterogeneity~\cite{stauffer2014introduction,stauffer1979scaling,newman2001random,goltsev2003critical}. For
instance, in square lattices, $p_c\approx 0.5927$, and in any
two-dimensional lattice, $\beta=5/36$ and
$\tau=187/91$~\cite{stauffer2014introduction}. On the other hand, it
was shown that for an uncorrelated random network with a degree
distribution $P(k)$ (i.e., the fraction of nodes with connectivity or
degree $k$), the threshold $p_c$ is given by $p_c=1/\left( \langle k^2
\rangle/\langle k\rangle-1 \right)$~\cite{newman2001random}, where
$\langle k \rangle$ and $\langle k^2\rangle$ are the first- and
second-order moments of $P(k)$,
respectively~\cite{cohen2000resilience}. In random homogeneous
networks (i.e. with $\langle k^2 \rangle <\infty$) the percolation
threshold is finite, such as in Erd\H{o}s-R\'enyi networks (ER) with a
Poisson degree distribution $P(k)=\langle k \rangle^k\exp(-\langle k
\rangle)/k!$. Furthermore, in these types of networks, $\beta=1$ and
$\tau=5/2$~\cite{newman2001random}. On the other hand, for
heterogeneous networks (i.e., with $\langle k^2\rangle=\infty$), such
as scale-free (SF) networks with a degree distribution $P(k)\sim
k^{-\lambda}$ with $2<\lambda<3$, the percolation threshold is zero,
implying that the GC is very robust to random node failures. Moreover,
using Tauberian theorems, Ref.~\cite{cohen2002percolation} proved that
$\beta=|3-\lambda|^{-1}$ and $\tau=2+(\lambda-2)^{-1}$ for
$2<\lambda<4$, and the percolation transition is higher than
second-order. Recently, Radicchi and Castellano~\cite{radicchi2015breaking} showed that for random
link and node percolation in SF networks with $\lambda\le 3$, the
fraction of nodes that belong to the GC behaves above $p=0$ as $P_{\infty}\sim
p^{\beta_b}$ and $P_{\infty}\sim p^{\beta_s}$,
respectively, in which the exponents $\beta_s$ and $\beta_b$ are related by
$\beta_s=\beta_b+1$. Other works have also studied a node percolation
(and link percolation) model in random networks with community
structure and its critical
exponents~\cite{dong2018resilience,valdez2018role,ma2020role}.

Besides node percolation, a wide range of percolation models have been
proposed to study different types of network failures, such as
targeted percolation~\cite{cohen2001breakdown}, l-hop
percolation~\cite{shang2011hop}, k-core
percolation~\cite{baxter2011heterogeneous}, and percolation in
interdependent networks~\cite{Buldyrev-2010}. Most of these models
consider that nodes can have only two mutually exclusive states
($n=2$): failed and non-failed. However, for some systems, it is more
realistic to include additional states in order to study the case
where nodes are partially damaged or have different
vulnerabilities. To investigate a percolation process in which the
number of mutually exclusive states is greater than two ($n>2$),
Krause et al.~\cite{krause2017color,krause2016hidden} developed a new
type of percolation model called "color-avoiding" percolation that is
useful for studying secured-message passing in communication
networks. In this model, nodes are separated into different classes or
colors, representing a shared vulnerability to failure. On the other
hand, Shang~\cite{shang2014vulnerability} proposed a percolation
process called "fractional percolation" with $n=3$. In this model,
nodes can be in one of the following mutually exclusive states: fully
functional (FF), partially functional (PF), and dysfunctional (D). FF
nodes can be connected to nodes in FF and PF states, while PF nodes
only have links to FF nodes, i.e. they lose their connections with
another PF node. This may represent a case in which two partially
damaged components in a wireless sensor or an electric network do not
have enough energy to communicate with each other, but they can
communicate with fully functional
components~\cite{moh2017dynamic,shang2014vulnerability}. In the
simplest version of this model, a fraction $1-q$ of nodes are FF,
while of the remaining $q$, a fraction $(1-r) q$ is PF and a fraction
$rq$ is D. Here, a giant component is defined as a macroscopic cluster
composed of FF and PF nodes. In that work, it was found that the
network undergoes a continuous phase transition and the structure is
more robust compared to random node percolation. However, the
geometrical structure of the network and the critical exponents around
the critical point for fractional percolation have not been studied
yet.

In this manuscript, we fill these gaps, finding that for a region in
the plane $r$ vs $q$, the topology can be described as a coarse-grained
bipartite network. In this region, the network is composed of finite
clusters of FF nodes that behave like
supernodes~\cite{kalisky2006scale} with a SF or power-law degree
distribution. Furthermore, we obtain that at $q=1-1/\left( \langle
k^2 \rangle/\langle k\rangle-1 \right)$, the fraction of FF and PF
nodes belonging to the GC decreases with $1-r$ as a power-law function
with exponents $\beta$ and $\beta^*=\beta+1$, respectively. For this
case, we show that the emergent coarse-grained bipartite network
explains the measured value of $\beta^*$.

Our manuscript is organized as follows: 1) in Sec.~\ref{Sec.Eq} we
present our equations for fractional percolation and compare our
theoretical solutions with those of
Ref.~\cite{shang2014vulnerability}, 2) in Sec.~\ref{Sec.qqc}, the critical exponents $\beta$ and $\beta^*$ are computed, 3) in
Sec.~\ref{app.taub} we study a bipartite network in order to explain the
 values of $\beta$ and $\beta^*$, and 4) in
Sec.~\ref{sec.conclu} we display our conclusions.

\section{Theoretical equations}\label{Sec.Eq}

In this section, we present the equations to compute $P_{\infty}$ and
$\langle s \rangle$ for fractional percolation,
using that the connections among FF and PF nodes form a semi-bipartite structure.

By definition, a bipartite network is composed of two groups of nodes
that we denote $A$ and $B$ (for instance, films and actors) in which
links only occur between nodes in different
groups~\cite{newman2001random}. The degree distribution of each group
is denoted as $P^A(k)$ and $P^B(k)$. Similarly, in fractional
percolation, there are two groups of functional nodes: FF and PF. In
turn, PF nodes cannot be connected to each other but only to FF
nodes. However, in contrast to a bipartite structure, an FF node can
be connected not only to PF nodes but also to FF nodes. In
consequence, this network structure is called
"semi-bipartite"~\cite{ciupala2016wave}. To investigate how the
network topology is affected by fractional percolation in the limit of
large network size ($N\to \infty$), we will use the generating
function formalism which describes the network structure as a
branching
process~\cite{wilf2005generatingfunctionology,newman2001random}. This
approach has been applied successfully in previous works to compute
different magnitudes, such as $P_{\infty}$ and $\langle s\rangle$ in
several percolation processes~\cite{Newman-2006,li2021percolation}. In
this approach, for a bipartite network, it is used: 1) the generating
function for the degree distribution of group $i=$\{$A,B$\},
$G_0^i[x]=\sum_{k=k_{min}}^{k_{max}}P^i(k) x^k$ where $k_{min}$ and
$k_{max}$ are the minimum and maximum degrees, and 2) the generating
function for the so-called excess degree distribution of group
$i=$\{$A,B$\}, $G_1^i[x]=\sum_{k=k_{min}}^{k_{max}}kP^i(k)/\langle k
\rangle x^{k-1}$. In a previous work on bipartite networks that used
the generating function formalism (see
Ref.~\cite{hooyberghs2010percolation}), it was shown that $P_{\infty}$
can be computed by solving two self-consistent equations, each
representing a branching process from one group to the
other. Nevertheless, for fractional percolation, we will need an
additional equation to consider a branching process between FF nodes,
i.e., nodes in the same group. In addition, it is important to note
that in fractional percolation, FF and PF nodes have the same degree
distribution because they are randomly selected from the same
substrate. Thus, FF and PF nodes have the same generating functions
for the degree distribution, which we denote as
$G_0[x]=\sum_{k=k_{min}}^{k_{max}} P(k) x^k$, that is, without any
superscript. Similarly, we denote $G_1[x]=\sum_{k=k_{min}}^{k_{max}} k
P(k)/\langle k\rangle x^{k-1}$ as the generating function for the
excess degree distribution for both FF and PF nodes.

The self-consistent equations for fractional percolation are:
\begin{eqnarray}
f_{FF\to FF}&=&1-G_1[qr+(1-q)(1-f_{FF\to FF})+q(1-r)(1-f_{FF\to PF})],\label{eq.fFF}\\
f_{FF\to PF}&=&1-G_1[qr+q(1-r)+(1-q)(1-f_{PF\to FF})],\label{eq.fFFPF}\\
f_{PF\to FF}&=&1-G_1[qr+(1-q)(1-f_{FF\to FF})+q(1-r)(1-f_{FF\to PF})], \label{eq.fPF}
\end{eqnarray}
where:
\begin{itemize}
\item $f_{FF\to FF}$ is the probability that a branching process from an FF node to an FF neighbor leads to the GC
\item $f_{FF\to PF}$ ($f_{PF\to FF}$) is the probability that a branching process from an FF (PF) node to a PF (FF) neighbor leads to the GC.
\end{itemize}
The second r.h.s. term of Eq.~(\ref{eq.fFF}) accounts for all
configurations in which an FF node reached through a link from an FF
node does not lead to the GC because its neighbors are either: 1)
dysfunctional with probability $qr$ (see Introduction), 2) fully
functional but they do not lead to the GC with probability
$(1-q)(1-f_{FF\to FF})$, or 3) partially functional but they do not
lead to the GC with probability $q(1-r)(1-f_{FF\to PF})$. On the other
hand, the second r.h.s term of Eq.~(\ref{eq.fFFPF}) considers all
configurations in which the neighbors of a PF node (reached through a
link from a FF node) do not lead to the GC because they
are either: 1) dysfunctional with probability $qr$, 2) PF with
probability $q(1-r)$, or 3) FF but they do not lead to the
GC with probability $(1-q)(1-f_{PF\to FF})$. Note that the r.h.s of
Eqs.~(\ref{eq.fFF}) and~(\ref{eq.fPF}) are equal, so $f_{FF\to
  FF}=f_{PF\to FF}$, and therefore the system of self-consistent
equations reduces to
\begin{eqnarray}
f_{FF\to FF}&=&1-G_1[qr+(1-q)(1-f_{FF\to FF})+q(1-r)(1-f_{FF\to PF})],\label{eq.fFF1old}\\
f_{FF\to PF}&=&1-G_1[qr+q(1-r)+(1-q)(1-f_{FF\to FF})].\label{eq.fFFPF1old}
\end{eqnarray}
Next, if we change the notation of $f_{FF\to FF}$ and $f_{FF\to PF}$
to $f_{FF}$ and $f_{PF}$, respectively, the above equations can be
rewritten as
\begin{eqnarray}
f_{FF}&=&1-G_1[qr+(1-q)(1-f_{FF})+q(1-r)(1-f_{PF})],\label{eq.fFF1}\\
f_{PF}&=&1-G_1[qr+q(1-r)+(1-q)(1-f_{FF})].\label{eq.fFFPF1}
\end{eqnarray}
where $f_{FF}$ ($f_{PF}$) can be interpreted simply as the probability
that an FF (PF) node reached by traversing a randomly chosen link
belongs to the GC.

After solving this system of equations, the fraction of nodes belonging to the GC can be computed as,
\begin{eqnarray}
  P_{\infty}=P_{\infty}^{FF}+P_{\infty}^{PF}, \label{eq.Pinf}
\end{eqnarray}
where the first and second r.h.s terms are the fractions of FF and PF
nodes belonging to the GC, respectively, which are described by the following equations
\begin{eqnarray}
P_{\infty}^{FF}&\equiv& (1-q)\left(1-G_0[qr+(1-q)(1-f_{FF})+q(1-r)(1-f_{PF})]\right),\label{eq.PinfFFF}\\
P_{\infty}^{PF}&\equiv& q(1-r)(1-G_0[q+(1-q)(1-f_{FF})]).\label{eq.PinfPF}
\end{eqnarray}

\begin{figure}[h]
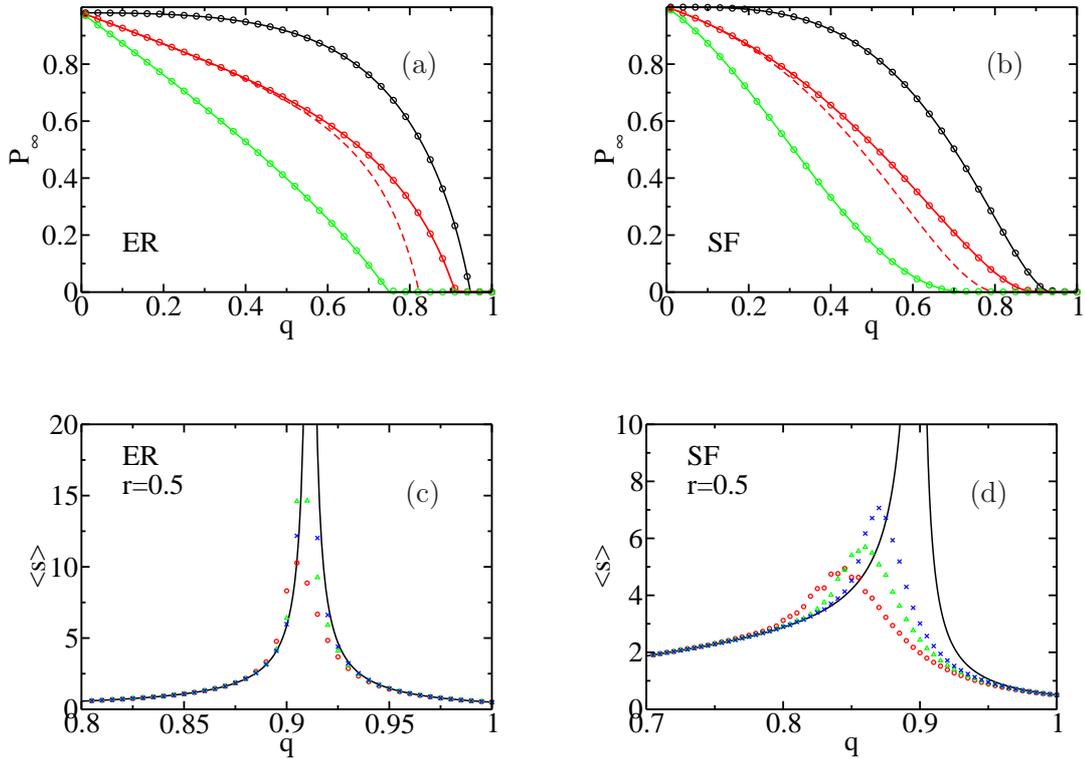

\vspace{0.5cm}
\begin{center}
\begin{overpic}[scale=0.25]{PinfER.eps}
  \put(80,55){(a)}
\end{overpic}
\vspace{1cm}
\hspace{1cm}
\begin{overpic}[scale=0.25]{PinfSF35.eps}
  \put(80,55){(b)}
\end{overpic}
\vspace{0.0cm}
\begin{overpic}[scale=0.25]{smedER.eps}
  \put(80,55){(c)}
\end{overpic}
\vspace{0.0cm}
\hspace{1cm}
\begin{overpic}[scale=0.25]{smedSF35.eps}
  \put(80,55){(d)}
\end{overpic}
\end{center}
\caption{Results for fractional percolation in ER and SF
  networks. Panel a: $P_{\infty}$ vs. $q$ for an ER network with
  $\langle k \rangle=4$ and several values of $r$: $r=0$ (black),
  $r=0.5$ (red), and $r=1$ (green). Symbols correspond to our
  stochastic simulations with $N=10^5$, the solid lines were obtained
  from Eq.~(\ref{eq.Pinf}), and the dashed lines were computed from
  the theory in Ref.~\cite{shang2014vulnerability}. Panel b:
  $P_{\infty}$ vs. $q$ for a SF network with $\lambda=3.5$ and
  $k_{min}=2$ (to ensure the existence of the percolation threshold
  for $\lambda>3$~\cite{newman2002random,cohen2000resilience}), and using the same
  parameter values as in panel a. Panel c: $\langle s\rangle$
  vs. $q$ for an ER network with $\langle k\rangle=4$ and $r=0.5$. The
  symbols correspond to our stochastic simulations for $N=10^5$ (red),
  $N=4\times 10^5$ (green), and $N=16\times 10^5$ (blue) and the solid
  line was obtained from Eq.~(\ref{eq.smed}). Panel d: $\langle
  s\rangle$ vs. $q$ for a SF network with $k_{min}=2$ and
  $\lambda=3.5$, and using the same parameters as in panel c. The
  stochastic results were obtained over $10^3$
  realizations.}\label{fig.FracPinfsmed}
\end{figure}

We compare the solution of Eq.~(\ref{eq.Pinf}) and the stochastic
simulations in Fig.~\ref{fig.FracPinfsmed}a-b for several values of
$r$ and for ER and SF networks with $\lambda=3.5$, obtaining an excellent agreement
between theory and simulations. In addition, we also include the
curves predicted by Ref.~\cite{shang2014vulnerability}, which do not
match the stochastic simulation results because the theory in
Ref.~\cite{shang2014vulnerability} uses only one self-consistent
equation and does not consider the semi-bipartite structure induced by
fractional percolation. The codes of our stochastic simulations
(written in Fortran 90) and the main equations are available at
GitHub~\cite{Exxon02}.

\begin{figure}[H]
\vspace{0.5cm}
\begin{center}
\begin{overpic}[scale=0.25]{HeatmapER.eps}
  \put(10,69){(a)}
\end{overpic}
\vspace{1cm}
\hspace{1cm}
\begin{overpic}[scale=0.25]{HeatmapSF.eps}
  \put(10,80){(b)}
\end{overpic}
\vspace{0.0cm}
\begin{overpic}[scale=0.75]{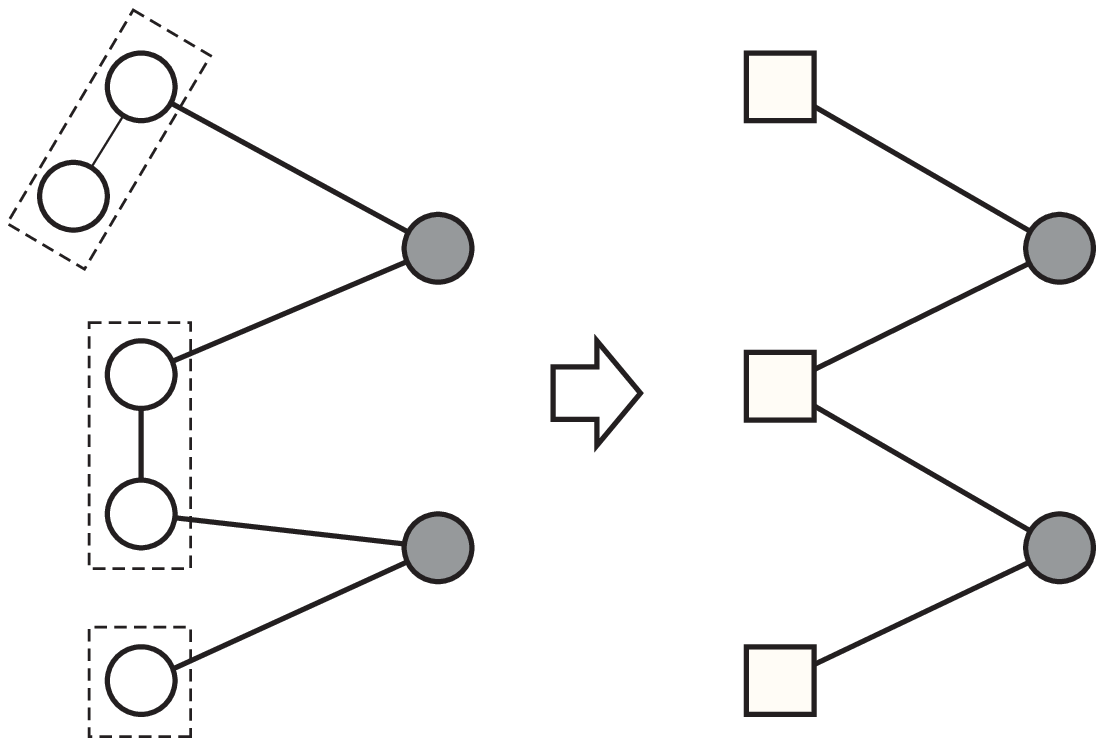}
  \put(0,70){(c)}
\end{overpic}
\vspace{0.0cm}
\end{center}
\caption{Heat-map of $P_{\infty}$ in the plane $q$-$r$ for an ER
  network with $\langle k \rangle=4$ (panel a) and for a SF network
  with $k_{min}=2$ and $\lambda=3.5$ (panel b), obtained from
  Eq.~(\ref{eq.Pinf}). The solid white line corresponds to $q_c(r)$
  computed from Eq.~(\ref{eq.qrcritic}), and the vertical dashed line
  indicates the value of $q_c(r=1)$. Panel c: On the left, we show an
  illustration of a functional cluster composed of FF nodes (white
  circles) and PF nodes (gray circles), in the regime $q_c(r=1)\leq
  q\leq q_c(r)$. Finite groups of FF nodes (FF-groups) are surrounded
  by dashed lines. On the right, we show the same cluster but where
  these FF-groups are replaced by supernodes
  (squares).}\label{fig.HeatmapPinf}
\end{figure}

Besides the relative size of the GC, another quantity of interest in
percolation theory is the average finite cluster size $\langle s
\rangle=\sum_{s=1}^{\infty}{s\;n_s}$ because it diverges when the
system undergoes a continuous phase transition at a critical point
(see Introduction). Following a similar reasoning to the derivation of
$P_{\infty}$, the average finite cluster size is given by
\begin{eqnarray}\label{eq.smed}
\langle s \rangle&=&(1-q)\frac{dH_{0,FF}[x]}{dx}\Bigr|_{x=1}+q(1-r)\frac{dH_{0,PF}[x]}{dx}\Bigr|_{x=1},
\end{eqnarray}
where $H_{0,FF}[x]$ and $H_{0,PF}[x]$ are the generating functions for
the cluster size distribution if a randomly chosen node is FF or PF,
respectively. The details to calculate $H_{0,FF}[x]$ and $H_{0,PF}[x]$
can be found in Appendix~\ref{app.H}. In
Fig.~\ref{fig.FracPinfsmed}c-b, we display $\langle s\rangle$ as a
function of $q$ for $r=0.5$ obtained from Eq.~(\ref{eq.smed}) and
simulations in ER and SF networks. As expected, we observe that the
theoretical solution diverges at a given value $q_c(r=0.5)$ for each
network topology, and the stochastic simulations converge to this
curve as the network size increases. In order to compute $q_c(r)$ for
any value of $r$, we use that at the critical point in a continuous
phase transition, the Jacobian matrix $J$ of the system of
Eqs.~(\ref{eq.fFF1})-(\ref{eq.fFFPF1}) satisfies
\begin{eqnarray}\label{eq.jacob}
\det(J-I)=0,
\end{eqnarray}
where $\det(\cdot)$ is the determinant function, $I$ is the identity
matrix, and $J$ is evaluated at $f_{FF}=f_{PF}=0$. After
straightforward calculations, we obtain that $q_c(r)$ is described by the following equation
\begin{eqnarray}\label{eq.qrcritic}
  \frac{(G_1^{'}[1])^2(1-q_c)q_c-(1-G_1^{'}[1](1-q_c))}{(G_1^{'}[1])^2(1-q_c)q_c}=r,
\end{eqnarray}
where $G_1^{'}[x]\equiv dG_1[x]/dx$. In particular, for $r=1$, that
is, when there are only FF and D nodes, we recover that
$q_c(r=1)=1-1/G_1^{'}[1]=1-1/(\langle k^2\rangle/\langle k\rangle -1)$
which is the critical threshold for random node percolation.  In
Figs.~\ref{fig.HeatmapPinf}a-b, we show the heat-map of $P_{\infty}$
in the plane $q-r$ for ER and SF networks obtained from
Eq.~(\ref{eq.Pinf}). In addition, we also include in these figures,
the critical line $q_c(r)$ predicted by Eq.~(\ref{eq.qrcritic}) (solid
white line), and the vertical line which indicates the value of
$q_c(r=1)$ (dashed white line). On the right side of the critical
line, the network is composed only of finite clusters, whereas on the
left side, a GC of functional nodes (FF and PF) emerges. However, it
is important to note that there are two regimes in this percolating
phase. For $q_c(r=1)\leq q \leq q_c(r)$, i.e., in the region between
the dashed and solid lines, FF nodes alone cannot form a GC because
when $r=1$ (i.e., when PF nodes are absent),
$P_{\infty}=0$. Therefore in this regime, the GC structure is
composed of finite groups of FF nodes, that we call FF-groups, which
are connected by PF nodes, as shown in the schematic illustration in
Fig.~\ref{fig.HeatmapPinf}c. In consequence, this structure can be
considered as a coarse-grained bipartite network in which one group is
composed of PF nodes, and the other group by FF-groups~\footnote{Note
  that this region does not exist for SF networks with $\lambda\le 3$
  because $q_c(r=1)=1$. Moreover, the network is composed of only D
  and PF nodes at $q_c(r=1)=1$.}. On the other
hand, on the left side of the dashed line ($q<q_c(r=1)$),
the GC emerges regardless of the value of $r$ because even in the
extreme case of $r=1$ when all PF nodes are removed, there are enough
FF nodes to form this GC.

In the following sections, we will study how the structure induced by
fractional percolation in the percolating phase impacts the critical
behavior of several magnitudes at $q_c(r=1)$.

\section{Critical behavior at $q=q_c(r=1)$ in random networks}\label{Sec.qqc}
In this section, we will investigate the asymptotic behavior of the
relative size of the GC when $r$ approaches $r=1$ with
$q=q_c(r=1)$ in random
networks. We recall that $r$ only controls the total fraction of PF
and D nodes but not the total fraction of FF nodes which remains fixed
for a given value of $q$.

\begin{figure}[H]
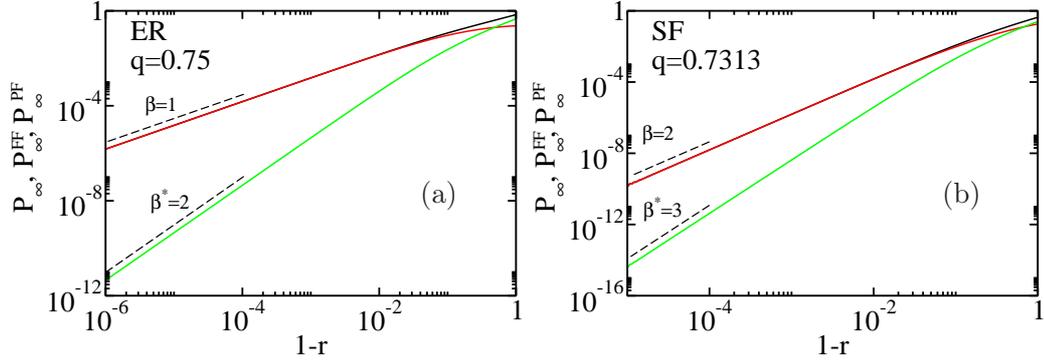

\vspace{0.5cm}
\begin{center}
\begin{overpic}[scale=0.25]{CritPinfER.eps}
  \put(80,30){(a)}
\end{overpic}
\vspace{0.5cm}
\begin{overpic}[scale=0.25]{CritPinfSF.eps}
  \put(80,30){(b)}
\end{overpic}
\vspace{0.0cm}
\end{center}
\caption{$P_{\infty}$ (black solid line),
  $P_{\infty}^{FF}$ (red solid line), and $P_{\infty}^{PF}$ (green
  solid line) as functions of $1-r$ for: an ER network with $\langle
  k\rangle =4$ and $q=q_c(r=1)=0.75$ (panel a), and a SF network with
  $\lambda=3.5$ and $k_{min}=2$ and $q=q_c(r=1)\approx 0.7313$ (panel
  b) in a log-log scale. Solid lines were obtained from
  Eqs.~(\ref{eq.Pinf})-(\ref{eq.PinfPF}), and dashed lines represent a
  power-law fit of $P_{\infty}^{FF}$ and $P_{\infty}^{PF}$.}\label{fig.CritPinf}
\end{figure}

In Figs.~\ref{fig.CritPinf}a and b, we show $P_{\infty}$ and $P_{\infty}^{FF}$ as functions
of $1-r$ for an ER network with $\langle k\rangle=4$ at
$q=q_c(r=1)=1-1/\langle k\rangle$, and for a SF network with
$\lambda=3.5$, respectively. We obtain that $P_{\infty}$ and $P_{\infty}^{FF}$  scale as a
power-law with exponent $\beta=1$ (for the ER network) and $\beta=2$
(for the SF network) as in random node percolation (see
Introduction). On the other hand, Figs.~\ref{fig.CritPinf}a and b show that
$P_{\infty}^{PF}\sim(1-r)^{\beta^*}$ with $\beta^*=2$ (for the ER
network) and $\beta^*=3$ (for the SF network), which differ from the
exponents  for $P_{\infty}$ and $P_{\infty}^{FF}$.

\begin{figure}[H]
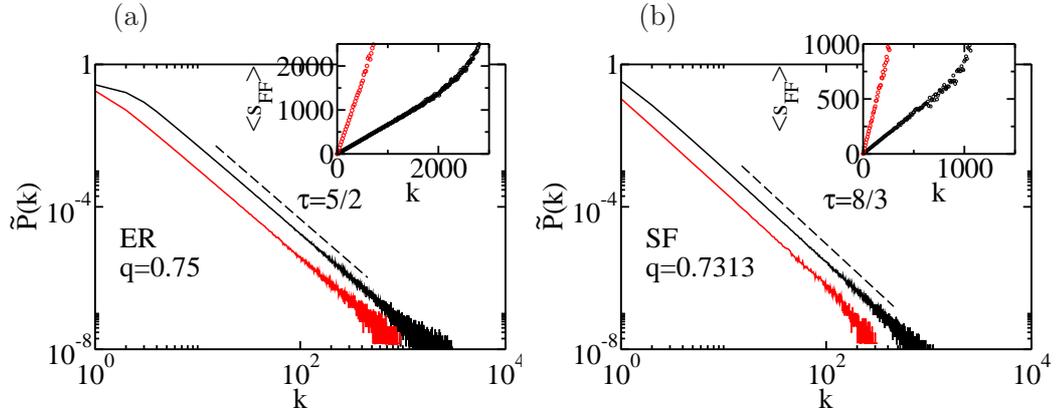

  \vspace{0.5cm}
  \begin{center}
\begin{overpic}[scale=0.25]{PkeffER.eps}
  \put(20,75){(a)}
\end{overpic}
\vspace{0.0cm}
\begin{overpic}[scale=0.25]{PkeffSF.eps}
  \put(20,75){(b)}
\end{overpic}
\vspace{0.0cm}
\end{center}
\caption{Degree distribution of FF-groups, $\tilde{P}(k)$, at $q=q_c$
  for $r=0.50$ (black solid line) and $r=0.90$ (red solid line) for:
  an ER network with $\langle k\rangle =4$ and $q=q_c(r=1)=0.75$
  (panel a), and a SF network with $\lambda=3.5$ and $k_{min}=2$ and
  $q=q_c(r=1)\approx 0.7313$ (panel b) in a log-log scale. The dashed
  lines are power-law functions with exponent $\tau$ (see
  Introduction). Insets: Average mass of an FF-group, $\langle s_{FF}\rangle$, as a function of
  its degree for $r=0.50$ (black) and $r=0.90$ (red) displayed in a
  double linear scale. Simulations have been averaged over 1000
  network realizations with $N=1.6\times 10^6$.}\label{fig.PkeffMass}
\end{figure}

In Fig.~\ref{fig.PkeffMass}a and b, from our stochastic simulations at $q=q_c(r=1)$ and
different values of $r$, we observe that the set of FF-groups
behaves as a set of ``supernodes'' (following similar reasoning as in
Ref.~\cite{kalisky2006scale}) with a power-law degree distribution,
$\tilde{P}(k)\sim k^{-\tau}$,
where:
\begin{enumerate}
\item we define the degree of an FF-group as the number $k$ of PF
  nodes connected to this group,
\item $\tau\in(2,3)$ is the exponent for the
distribution of finite cluster sizes in random node percolation (see
Introduction).
\end{enumerate}
Therefore, we obtain that at $q=q_c(r=1)$, a bipartite
network emerges in which ``supernodes'' follow a heterogeneous degree
distribution. We also observe from our simulations that the average mass of an FF-group,
$\langle s_{FF} \rangle$, is a linear function of its degree (see insets in
Fig.~\ref{fig.PkeffMass}).

In the following section, we show that the topology of this emergent bipartite
network explains the value of $\beta^*$ observed in
Figs.~\ref{fig.CritPinf}a-b.

\section{Critical behavior in a bipartite network}\label{app.taub}

In Sec.~\ref{Sec.Eq}, we observed that at $q=q_c(r=1)$, the GC for
fractional percolation is composed of finite FF-groups connected by PF
nodes. By merging these finite groups into supernodes (as shown in
Fig.~\ref{fig.HeatmapPinf}c), we observed that a bipartite network
emerges. Moreover, from our simulations, we obtained that the:
\begin{itemize}
\item FF-groups have a degree distribution with a power-law tail with
  exponent $\tau\in \left(2,3\right)$, where $\tau=5/2$ for ER
  networks, and $\tau=2+(\lambda-2)^{-1}$ for SF networks.
\item PF nodes have a diluted degree distribution: 1) which is
  homogeneous if the substrate network is homogeneous, or 2) has a
  power-law tail with exponent $\lambda$ if the substrate network is a
  SF network with exponent $\lambda$ (not shown here).
\end{itemize}.
Finally, we obtained from our theoretical equations that the fraction
of PF nodes belonging to the GC, $P_{\infty}^{PF}$, decays faster than
$P_{\infty}$ and $P_{\infty}^{FF}$ in the limit of $r\to 1$, and it follows a
power-law function with exponent $\beta^*$.

In the following, we will introduce and study a simplified model which
captures the main topological features of the emergent network in
order to explain the value of the exponent $\beta^*$ for an
ER network (see Fig.~\ref{fig.CritPinf}a). Specifically, we will
consider a bipartite network in which:
\begin{itemize}
\item one group, called ``sf'', has a
pure power-law degree distribution given by $P^{sf}(k)=\epsilon
k^{-\tau}$ ($\epsilon$ is a normalization constant) with $\tau=5/2$,
\item  the other group, called ``h'', has a homogeneous degree
distribution $P^{h}(k)$, i.e., the second-order moment of $P^{h}(k)$
does not diverge.
\end{itemize}

The generating functions for the degree
distributions if a randomly node is chosen, are:
\begin{itemize}
\item for group ``sf'': $G_{0,sf}[x]=\sum_{k=1}^{\infty}
  \epsilon k^{-\tau}x^k=Li_{\tau}[x]/Li_{\tau}[1]$ where $Li$ is the
  Polylogarithm function,
\item for group ``h'': $G_{0,h}[x]=\sum_{k=0}^{\infty} P^{h}(k)x^k$,
\end{itemize}
and the generating functions for the excess degree distributions if a
node is reached through a link, are given by:
\begin{itemize}
\item for group ``sf'': $G_{1,sf}[x]=\sum_{k=1}^{\infty} k
  \epsilon k^{-\tau}x^{k-1}/\langle
  k^{sf}\rangle=x^{-1}Li_{\tau-1}[x]/Li_{\tau-1}[1]$ where $\langle
  k^{sf}\rangle\equiv \sum k \epsilon k^{-\tau}$,
\item for group ``h'': $G_{1,h}[x]=\sum_{k=1}^{\infty} k
  P^{h}(k)x^{k-1}/\langle k^h\rangle$ where $\langle k^h\rangle=\sum k
   P^h(k)$.
\end{itemize}

Analogously to fractional percolation in which a fraction $r$ of PF
nodes were removed at $q=q_c(r=1)$ whereas FF-groups remained intact
(see Sec.~\ref{Sec.qqc}), now we will study a percolation process in
our simplified model in which a fraction $r$ of nodes in group ``h''
are removed whereas nodes in group ``sf'' remain undamaged.

Following Ref.~\cite{newman2001random}, the self-consistent equations
to compute the GC are given by,
\begin{eqnarray}
  Q_{h}&=&r+(1-r)G_{1,h}[Q_{sf}],\label{eq.appQER}\\
  Q_{sf}&=&G_{1,sf}[Q_{h}],\label{eq.appQSF}
\end{eqnarray}
where $Q_{h}$ is the probability that a link from group ``sf''
to group ``h'' does not lead to the GC because: 1) the node in group
``h'' is removed with probability $r$, or 2) the node in group ``h''
is not removed but is not connected to the GC with probability
$(1-r)G_{1,h}[Q_{sf}]$. On the other hand, $Q_{sf}$ is the
probability that a node in group ``sf'' (reached through a link from
group ``h'') is not connected to the GC because none
of its outgoing links leads to the GC, with probability
$G_{1,sf}[Q_{h}]$.

Similarly to Eqs.~(\ref{eq.PinfFFF}) and~(\ref{eq.PinfPF}), we now define
$P_{\infty}^{h}$ and $P_{\infty}^{sf}$ as the fraction of nodes in
groups ``h'' and ``sf'' that belong to the GC~\footnote{These
  fractions are relative to the number of nodes in each group.},
respectively, which are described by the following equations:
\begin{eqnarray}
  P_{\infty}^h&=&(1-r)(1-G_{0,h}[Q_{sf}]),\label{eq.appPinfh}\\
  P_{\infty}^{sf}&=&1-G_{0,sf}[Q_{h}].\label{eq.appPinfsf}
\end{eqnarray}
where the r.h.s. of Eq.~(\ref{eq.appPinfh}) accounts for all
configurations in which a node in group ``h'' belongs to the GC
because: 1) is not removed with probability $1-r$, and 2) at least one
of its neighbors is connected to the GC with probability
$(1-G_{0,h}[Q_{sf}])$. Similarly, the r.h.s. of
Eq.~(\ref{eq.appPinfsf}) accounts for all configurations in which a
node in group ``sf'' belongs to the GC because at least one of its
neighbors is connected to this component with probability
$1-G_{0,sf}[Q_{h}]$. It is straightforward from
Eqs.(\ref{eq.appQER})-(\ref{eq.appPinfsf}) to see that $r=1$ is the
critical threshold for this network, i.e., all nodes in group ``h''
must be removed to destroy the GC in the limit of large network sizes
because the degree distribution of group ``sf'' is heterogeneous
($\tau<3$ as mentioned above).

To obtain the critical exponents of $P_{\infty}^h$ and $P_{\infty}^{sf}$
in the limit $r\to 1$, we expand the r.h.s of Eq.~(\ref{eq.appQER})
around $Q_{h}=1$ in a Taylor series keeping only terms below
first order, and for the r.h.s of Eq.~(\ref{eq.appQSF}) we apply
Tauberian theorems~\cite{durrett2007random,cohen2002percolation}
around $Q_{sf}=1$ using that $2<\tau<3$, yielding to the
following equations,
\begin{eqnarray}
  Q_{h}&\sim&r+(1-r)\left(1+\left(\frac{\langle (k^h)^2\rangle}{\langle k^h\rangle}-1\right)(Q_{sf}-1)\right),\\
  Q_{sf}&\sim&1-c_1(1-Q_{h})^{\tau-2},
\end{eqnarray}
where $c_1$ is a constant. The solutions of this system of equations
are,
\begin{eqnarray}
  Q_{h}&\sim&1-c_2(1-r)^{\frac{1}{3-\tau}},\label{eq.appQhsftau}\\
  Q_{sf}&\sim&1-c_3(1-r)^{\frac{(\tau-2)}{3-\tau}},\label{eq.appQsfhtau}
\end{eqnarray}
where $c_2$ and $c_3$ are positive constants.

Finally, after expanding the r.h.s of
Eqs.~(\ref{eq.appPinfh})-(\ref{eq.appPinfsf}) around $Q_{sf}=1$
and $Q_{h}=1$, and combining with
Eqs.~(\ref{eq.appQhsftau})-(\ref{eq.appQsfhtau}), we obtain
\begin{eqnarray}
  P_{\infty}^h&\sim&(1-r)^{\frac{1}{3-\tau}},\label{eq.pinfhexp}\\
  P_{\infty}^{sf}&\sim&(1-r)^{\frac{1}{3-\tau}}\label{eq.pinfsfexp}.
\end{eqnarray}

Therefore, for $\tau=5/2$ we get that the exponent of $P_{\infty}^h$
is $(3-\tau)^{-1}=2$ which is in agreement with the measured
exponent $\beta^*$ for $P_{\infty}^{PF}$ in
Fig.~\ref{fig.CritPinf}a. As a consequence, this result shows that the
emergent structure in fractional percolation explains the value of
$\beta^*$.

On the other hand, it would be expected that $P_{\infty}^{sf}$ and
$P_{\infty}^{FF}$ have the same critical exponent. However, by
comparing Eqs.~(\ref{eq.pinfhexp}) and~(\ref{eq.pinfsfexp}), it is
clear that the exponent for $P_{\infty}^{sf}$ is $(3-\tau)^{-1}=2$
which is different from the exponent $\beta=1$ for $P_{\infty}^{FF}$
(see Fig.~\ref{fig.CritPinf}a). To understand this discrepancy, it
should be noted that the size or mass of a node in group ``sf'' is
different from the mass of an FF-group. More specifically, when we
compute $P_{\infty}^{sf}$, we assume that the size of a single node in
group ``sf'' is equal to one. On the contrary, the mass of an FF-group
is proportional to its degree $k$ as shown in
Fig.~\ref{fig.PkeffMass}c. Therefore, based on the aforementioned
difference between masses, if we now assign a mass equal to $k$ to
each node with degree $k$ in the group ``sf'' and recalculate
$P_{\infty}^{sf}$ accordingly, we obtain (see details in
Appendix~\ref{app.enlar})
\begin{eqnarray}
  P_{\infty}^{sf}&\sim&(1-r)^{\frac{1}{3-\tau}-1}\label{eq.pinfsfexpNEW},
\end{eqnarray}
where for the homogeneous case ($\tau=5/2$) the exponent of
$P_{\infty}^{sf}$ is now $(3-\tau)^{-1}-1=1$ which is in agreement
with the measured exponent $\beta$ in
Fig.~\ref{fig.CritPinf}a. Finally, by comparing the exponents of
Eqs.~(\ref{eq.pinfsfexp}) and (\ref{eq.pinfsfexpNEW}), we obtain that
$\beta^*-1=\beta$.

Although the exponents in Eqs.~(\ref{eq.pinfhexp})
and~(\ref{eq.pinfsfexpNEW}) have been derived for the case where the
degree distribution of group ``h'' is homogeneous (with $\tau=5/2$), if
we repeat the same calculations for the following bipartite network:
\begin{itemize}
\item  the group ``h'' has a power-law degree distribution with
$3<\lambda<4$ ,
\item the group ``sf'' has a
pure power-law degree distribution given by $P^{sf}(k)=\epsilon
k^{-\tau}$ ($\epsilon$ is a normalization constant) with $\tau=2+(\lambda-2)^{-1}$,
\end{itemize}
we obtain that Eqs.~(\ref{eq.pinfhexp}) and~(\ref{eq.pinfsfexpNEW}) still
hold, and $\beta^*-1=\beta$. In particular, for $\lambda=3.5$, we get
that $\beta^*=(3-\tau)^{-1}=(\lambda-2)/(\lambda-3)=3$, and $\beta=2$
which are in agreement with the measured exponents $\beta^*$ and
$\beta$ in Fig.~\ref{fig.CritPinf}b.

In Appendix~\ref{app.more}, we show that the relation between $\beta$
and $\beta^*$ also holds for fractional percolation in SF networks
with $\lambda=3.25$, and $\lambda=3.75$, and in square lattices.

\section{Summary and Conclusion}\label{sec.conclu}

In this manuscript, we study a fractional percolation model in complex
networks. We find the exact equations governing the size of the GC and
the average size of finite clusters in the limit of large network
sizes, using that the connections among FF and PF nodes form a
semi-bipartite structure. Moreover, in the $q-r$ plane, we obtained
two functional regimes: one in which the GC does not need PF nodes to
emerge and another in which the GC is composed of finite groups of FF
nodes (FF-groups) that are connected by PF nodes. In the latter
regime, the GC can be described as a coarse-grained bipartite network,
and at $q=q_c(r=1)$, the FF-groups behave as a set of supernodes with
a power-law degree distribution. We also find that at $q=q_c(r=1)$ and
in the limit of $r\to 1$, the fraction of FF and PF nodes behave as
power-law functions of $1-r$ with exponents $\beta$ and
$\beta^*=\beta+1$, respectively, where the value of $\beta$ is the
same as in random node percolation. Furthermore, we show that the
value of $\beta^*$ can be explained by the emergent bipartite network
in which supernodes have a power-law degree distribution. Our present
findings could be easily extended to consider more node states ($n>3$), and
also to the case in which $q$ and $r$ depend on the node's degree. Our
work could also be extended to other percolation processes, such as
k-core and bootstrap
percolation~\cite{di2019insights,baxter2010bootstrap,dorogovtsev2006k}.
In addition, it would be interesting to characterize the network
structure further using recently developed tools and concepts, such as
the degree-degree correlation and the spectra of the
GC~\cite{tishby2018revealing}, as well as articulation points and
"bredges"~\cite{tishby2018statistical,bonneau2020statistical}. These
research directions are left for future work.

\section{Acknowledgments}
The authors acknowledge UNMdP (EXA 956/20), and CONICET for financial
support.

\appendix

\section{Generating functions for finite cluster sizes, and $\langle s\rangle$}\label{app.H}
Let $H_0[x]=\sum_{s=0}^{\infty} P(s)x^s$ be the generating function
for finite cluster sizes when a node is randomly chosen, where $P(s)$
is the probability that the chosen cluster has size $s$. This function
is given by,
\begin{eqnarray}\label{eq.appH0}
H_0[x]&=&qr+(1-q)H_{0,FF}[x]+q(1-r)H_{0,PF}[x],
\end{eqnarray}
where the first term corresponds to dysfunctional nodes (i.e. clusters
with size $s=0$), the second and third terms correspond to the
cases when the chosen node is FF and PF, respectively, and the
generating functions $H_{0,FF}[x]$ and $H_{0,PF}[x]$ are given by,
\begin{eqnarray}
  H_{0,FF}[x]&=&xG_0[qr+(1-q)H_{1,FF}[x]+q(1-r)H_{1,PF}[x]],\label{eq.appH0FF}\\
  H_{0,PF}[x]&=&xG_0[qr+q(1-r)+(1-q)H_{1,FF}[x]].\label{eq.appH0PF}
\end{eqnarray}
In Eq.~(\ref{eq.appH0FF}), the r.h.s accounts for all configurations
in which a FF node belongs to a finite cluster because its neighbors
are either: 1) D with probability $qr$, 2) FF (PF) and they lead only to a
finite cluster of functional nodes with size $s$ whose generating function is given
by $H_{1,FF}[x]$ ($H_{1,PF}[x]$). On the other hand, Eq.~(\ref{eq.appH0PF}) has a similar
interpretation as Eq.~(\ref{eq.appH0FF}). Following
Refs.~\cite{newman2001random,meyers2006predicting}, the generating functions $H_{1,FF}[x]$ and
$H_{1,PF}[x]$ are given by the following self-consistent functional
equations,
\begin{eqnarray}
  H_{1,FF}[x]&=&xG_1[qr+(1-q)H_{1,FF}[x]+q(1-r)H_{1,PF}[x]],\\
  H_{1,PF}[x]&=&xG_1[qr+q(1-r)+(1-q)H_{1,FF}[x]],\label{eq.appH1PF}
\end{eqnarray}
which have a similar interpretation as
Eqs.~(\ref{eq.fFF1})-(\ref{eq.fFFPF1}). Finally, after solving
Eqs.~(\ref{eq.appH0})-(\ref{eq.appH1PF}), we compute the mean finite cluster size as
\begin{eqnarray}
\langle s \rangle =\frac{dH_0}{dx}\Bigr|_{x=1}&=&(1-q)\frac{dH_{0,FF}}{dx}\Bigr|_{x=1}+q(1-r)\frac{dH_{0,PF}}{dx}\Bigr|_{x=1}.
\end{eqnarray}

\section{Enlarging a bipartite network}\label{app.enlar}
In Sec.~\ref{app.taub}, we obtained that for a bipartite network, the
fraction of nodes in group ``sf'' that belong to the GC obeys
Eq.~(\ref{eq.appPinfsf}), i.e.,
\begin{eqnarray}
P_{\infty}^{sf}=1-\sum_{k=1}^{\infty}P^{sf}(k)\left(Q_{h}\right)^k,\label{eq.ap.Psumg0}
\end{eqnarray}
where $P^{sf}(k)=\epsilon k^{-\tau}$ (with $2<\tau<3$), and $\epsilon$ is a normalization
constant. Moreover, we found that in the limit of $r\to 1$:
\begin{eqnarray}
P_{\infty}^{sf}\sim (1-r)^{\frac{1}{3-\tau}}\sim (1-r)^{\beta^*}.\label{eq.ap.betstar}
\end{eqnarray}
As mentioned in Sec.~\ref{app.taub}, in order to compute
$P_{\infty}^{sf}$, we assumed that the mass of a single node in group
``sf'' is equal to one. On the contrary, when we compute
$P_{\infty}^{FF}$, we observed that the mass of a single FF-group of
connectivity $k$ is proportional to its degree (see
Fig.~\ref{fig.PkeffMass}c).

In the following, we recalculate $P_{\infty}^{sf}$ for the case in
which each node with degree $k$ has a mass equal to $k$,
i.e., we will ``enlarge'' the group ``sf''~\footnote{Qualitatively, this
transformation can be pictured as the reverse process illustrated in
Fig.~\ref{fig.HeatmapPinf}c, that is, we replace each node in group ``sf'' with
degree $k$ by a finite FF-group whose mass is proportional to
$k$.}. More specifically, we have
to replace the distribution $P^{sf}(k)$ in Eq.~(\ref{eq.ap.Psumg0}) by
$kP^{sf}(k)/\langle k^{sf}\rangle$, where $\langle k^{sf}\rangle=\sum
kP^{sf}(k)$ is a normalization constant. By doing this,
Eq.~(\ref{eq.ap.Psumg0}) becomes,
\begin{eqnarray}
  P_{\infty}^{sf}&=&1-\sum_{k=1}^{\infty}\frac{kP^{sf}(k)}{\langle k^{sf} \rangle}\left(Q_{h}\right)^k,\label{eq.ap.Psumg1}\\
  &=&1-\sum_{k=1}^{\infty}\frac{\epsilon k^{-\tau+1}}{\langle k^{sf} \rangle}\left(Q_{h}\right)^k=1-Q_{h}\sum_{k=1}^{\infty}\frac{\epsilon k^{-\tau+1}}{\langle k^{sf} \rangle}\left(Q_{h}\right)^{k-1},\\
  &=&1-Q_{h}G_{1,sf}[Q_{h}],\label{eq.ap.Psumg1c}
\end{eqnarray}
where $G_{1,sf}[x]$ is the generating function for the excess degree
distribution of a node in group ``sf'' (see Sec.~\ref{app.taub}).

Using Tauberian theorems in the limit of $Q_{h}\to 1$ (see Sec.
4.3 in~\cite{durrett2007random}) , we have that the generating
function $G_{1,sf}[x]$ behaves as
\begin{eqnarray}\label{g1Taub}
  G_{1,sf}[x]\sim 1-c(1-x)^{\tau-2},
\end{eqnarray}
where $c$ is a constant, and we use the symbol $x\sim y$ to mean that
$x/y\to 1$~\cite{FellerVol2}. Then, if we combine
Eqs.~(\ref{eq.ap.Psumg1c}) and~(\ref{g1Taub}), we obtain that
$P_{\infty}^{sf}$ can be rewritten as,
\begin{eqnarray}
  P_{\infty}^{sf}&\sim&1-\left(1-c(1-Q_{h})^{\tau-2}\right),\\
  &\sim &c\left( 1-Q_{h}\right)^{\tau-2},\\
  &\sim&c(1-r)^{\frac{\tau-2}{3-\tau}},\\
  &\sim&c(1-r)^{\beta^*-1},\label{eq.appbetnew}
\end{eqnarray}
where in the last two steps we have recalled Eq.~(\ref{eq.appQhsftau})
and used $\beta^*=1/(3-\tau)$ (see Sec.~\ref{app.taub}). Therefore,
after comparing Eqs.~(\ref{eq.ap.betstar}) and~(\ref{eq.appbetnew}),
we obtain that enlarging the network size has changed the critical
exponent of $P_{\infty}^{sf}$ from $\beta^*$ to $\beta^*-1$.

\section{Critical exponents for SF networks and square lattices}\label{app.more}
\begin{figure}[H]
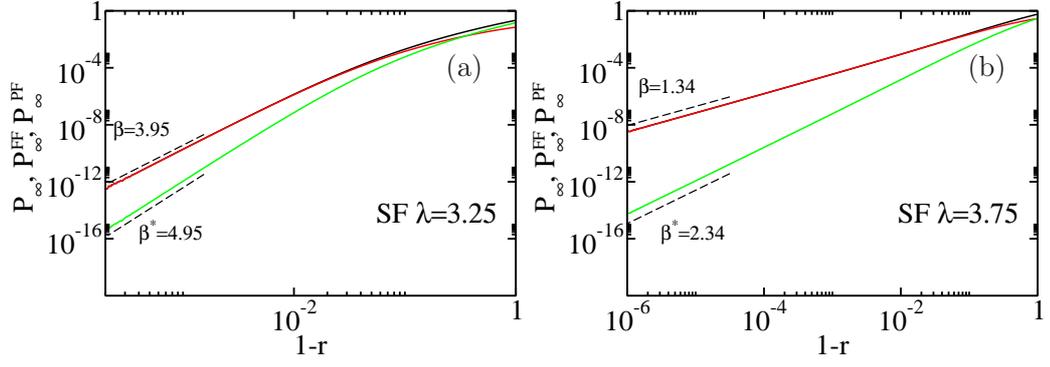

\vspace{0.5cm}
\begin{center}
\begin{overpic}[scale=0.25]{CritPinfSF325.eps}
  \put(85,55){(a)}
\end{overpic}
\vspace{0.5cm}
\begin{overpic}[scale=0.25]{CritPinfSF375.eps}
  \put(85,55){(b)}
\end{overpic}
\end{center}
\caption{Log-log plot of $P_{\infty}$ (solid black line),
  $P_{\infty}^{FF}$ (solid red line), and $P_{\infty}^{PF}$ (solid
  green line) as functions of $1-r$ at $q=q_c(r=1)$ for SF networks
  with $\lambda=3.25$ (panel a) and $\lambda=3.75$ (panel b). Solid
  lines were obtained from Eqs.~(\ref{eq.Pinf})-(\ref{eq.PinfPF}) and
  dashed lines represent a power-law fit. }\label{fig.moreSF}
\end{figure}

In Fig.~\ref{fig.moreSF}, we show for fractional percolation:
$P_{\infty}$, $P_{\infty}^{FF}$, and $P_{\infty}^{PF}$ as functions of
$1-r$ at $q=q_c(r=1)$ for SF networks with $\lambda=3.25$ (panel a)
and $\lambda=3.75$ (panel b), obtained from
Eqs.~(\ref{eq.Pinf})-(\ref{eq.PinfPF}). From these figures, we can see
that the measured values of the exponents $\beta$ and $\beta^*$ are in
agreement with the theoretical ones obtained in Sec.~\ref{app.taub},
i.e., $\beta^*=1/(3-\tau)$ with $\tau=2+(\lambda-2)^{-1}$ (see
Introduction), and $\beta=\beta^*-1=1/(\lambda-3)$ which corresponds
to the value of $\beta$ in random node percolation (see Introduction).

\begin{figure}[H]
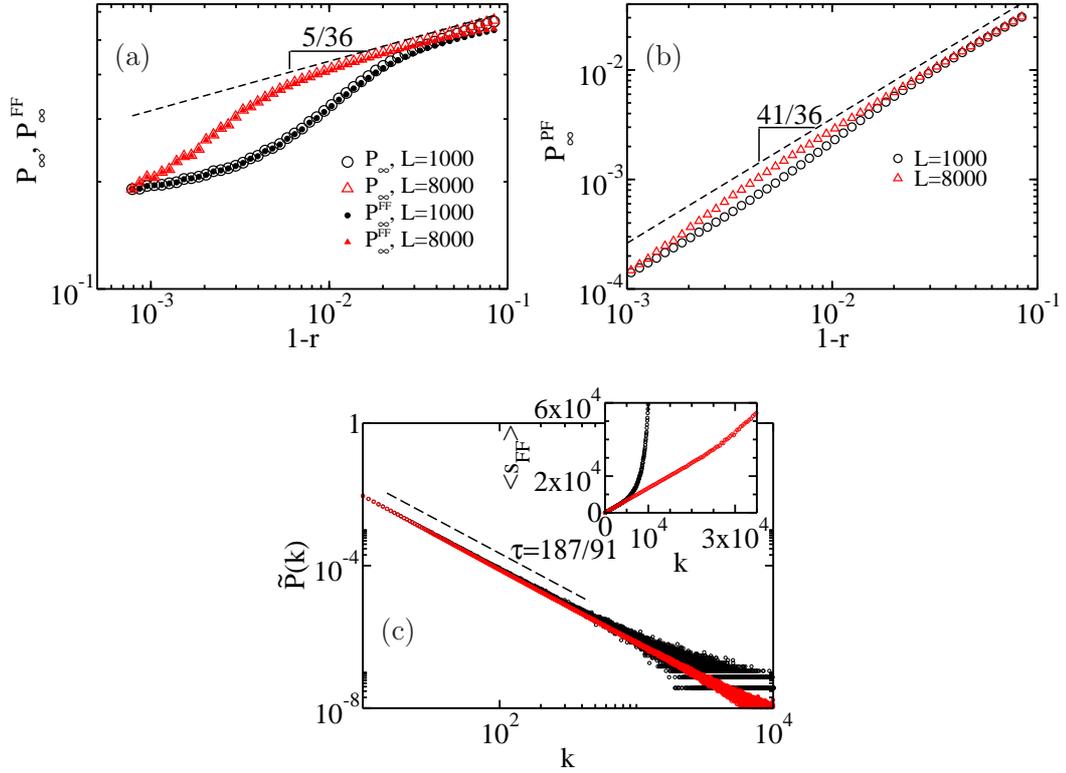

\vspace{0.5cm}
\begin{center}
\begin{overpic}[scale=0.25]{PinTotFF_Square.eps}
  \put(20,55){(a)}
\end{overpic}
\vspace{0.5cm}
\begin{overpic}[scale=0.25]{PinfPF_SquareNew.eps}
  \put(20,55){(b)}
\end{overpic}
\vspace{0.5cm}
\begin{overpic}[scale=0.25]{Pkeff2D.eps}
  \put(20,25){(c)}
\end{overpic}
\end{center}
\caption{Panel a: Log-log plot of $P_{\infty}$ (empty symbols) and
  $P_{\infty}^{FF}$ (full symbols) as functions of $1-r$ at
  $q=q_c(r=1)\approx 0.41$ for a square lattice of size $L\times L$
  with $L=1000$ (black circles), and $L=8000$ (red triangles), and
  rigid boundary conditions. Symbols are simulations averaged over 500
  realizations, and the dashed line corresponds to a power-law
  function with exponent $5/36$. Panel b: $P_{\infty}^{PF}$ for the
  same lattice in panel a, for $L=1000$ (black circles) and $L=8000$
  (red triangles). The dashed line corresponds to a power-law function
  with exponent $5/36+1=41/36$. Panel c: Degree distribution of FF-groups $\tilde{P}(k)$ at $q=q_c$
  for $r=0.50$, $L=1000$ (black symbols) and $L=8000$ (red symbols) in a log-log scale. The dashed
  line is a power-law function with exponent $\tau$ (see
  Introduction). In the inset we show the average mass of an FF-group, $\langle s_{FF}\rangle$,
  as a function of its degree for $r=0.50$, and $L=1000$ (black) and
  $L=8000$ (red) displayed in a double linear scale. Simulations have been averaged over 1000 network
  realizations.  }\label{fig.Square}
\end{figure}

Similarly, for square lattices (see Fig.~\ref{fig.Square}), we observe
that simulations support the relation $\beta^*=\beta+1=41/36$, where
$\beta=5/36$ (see Introduction). In addition, we also obtain from our simulations that the
degree distribution of FF-groups is a power-law function with exponent $\tau$
and the average mass of an FF-group is proportional to its degree (see
Fig.~\ref{fig.Square}c).

\bibliography{bib}

\end{document}